\documentstyle[12pt]{article}
\renewcommand{\theequation}{\arabic{section}.\arabic{equation}}

\begin{document}

\title{Quantum mechanics on Riemannian Manifold in Schwinger's
Quantization Approach I}

\author{Chepilko Nicolai Mikhailovich\\
\small\it
Physics Institute of the Ukrainian Academy of Sciences, 
Kyiv-03 028, Ukraine \\ \small\it e-mail: chepilko@zeos.net
\and
Romanenko Alexander Victorovich\\
\small\it
Kyiv Taras Shevchenko University,
Department of Physics, Kyiv-03 022, Ukraine\\
\small\it e-mail: ar@ups.kiev.ua}

\maketitle

\begin{abstract}
Schwinger's quantization scheme is extended in order to solve the
problem of the formulation of quantum mechanics on a space with a group
structure. The importance of Killing vectors in a quantization scheme
is showed.  Usage of these vectors provides algebraic properties
of operators to be consistent with the geometrical structure of a
manifold. The procedure of the definition of the quantum Lagrangian of
a free particle and the norm of velocity (momentum) operators is
given. These constructions are invariant under a general coordinate
transformation. The unified procedure for constructing the quantum
theory on a space with a group structure is developed. Using it quantum
mechanics on a Riemannian manifold with a simply transitive group
acting on it is investigated.

\end{abstract}


\section{Introduction}

There are many works \cite{sugano}-\cite{tsutsui} where
authors have represented the formulations of quantum mechanics on
spaces with a group structure. These investigations have been
carried out not only for the sake of academician interest, but
also for particular applications. The development of quantum mechanics
on curved spaces has been essentially stimulated by fruitful
investigations of several non-linear models such as the quantized
Skyrme model of simplest baryons \cite{skyrme1} - \cite{skyrme4},
the theory of three dimensional quantum chiral solitons
\cite{chiral1},\cite{chiral2} and supersymmetrical \cite{susy}
fermi-solitons. The main interest in these works has been caused
by the problem of determination of a so-called ``quantum
potential'' and the gauge structure of quantum mechanics on a curved
space (or a surface embedded into the Euclidean space). An
operator ordering problem in the starting Lagrangian has not been
exhaustively analyzed in these works. Moreover, some assumptions
(which seem to be truthful) are used in explicit or hidden form
without any rigorous definition.

In this connection we try to examine the complex of problems
associated with quantum mechanics on a curved space equipped with
a Riemannian metric and to give (as it possible) the rigorous
motivation of it. Our version of the formulation of quantum
mechanics on a curved space is based on Schwinger's action
principle \cite{schwinger}. Developing a quantization procedure,
we show that permissible variations (appearing in the action
principle) on such a manifold coincide with Killing vectors which
represent the group of isometries for the Riemannian metric.
Hence, the quantum-mechanical properties of the theory (an
operator algebra and a gauge structure) are determined by this
group.

The use of this procedure enable one to construct the quantum
Lagrangian by a rigorous way and to treat a ``quantum potential''
as a correction in the quantum Lagrangian which makes it a quantum
scalar under a general coordinate transformation. We also
introduce the definition of the scalar norm of momentum and
velocity operators which plays an important role in constructing
the quantum Lagrangian.

Our results, obtained by means of this procedure, are represented
in three papers. In the present work (first of them) we carry out
the main principles and extension of Schwinger's quantization
procedure for the case of a Riemannian space which is applicated
in order to construct quantum mechanics on a manifold with the simply
transitive group of isometries acting on it. Here Killing vectors
form the representation of the Lie algebra and its number is equal
to the dimension of a manifold. In this case Schwinger's quantization
procedure  is realized without any difficulties and obtained results
are in accordance with \cite{sugano}.

In the second paper we will consider the more complicated case of
quantum mechanics on a homogeneous Riemannian manifold $V_{n}$.
The number of Killing vectors is bigger than the dimension of the
manifold and they are not independent in a point $q_0\in V_{n}$.
Some of them form is the representation of the isotropy group of
$q_0\in V_{n}$. It turns out that quantum mechanics on the
homogeneous Riemannian manifold has a gauge structure and an
isotropy group acts as the group of local gauge transformations. A
gauge-fixing condition (which is necessary in this case) requires
a configuration space to be extended by adding new coordinates.
The quantum Lagrangian has to be modified by introducing new terms
(depending on new degrees of freedom) in such a way that local
gauge transformations become global ones. Obtained results are in
accordance with \cite{tsutsui}.

In the third work we will consider the formulation of quantum
mechanics on a Riemannian manifold with the intransitive group of
isometries. In this case the number of Killing vectors is fewer
than the dimension of a manifold. It will be shown that quantum
dynamics is completely determined only for degrees of freedom
which describe the invariant subspace (associated with Killing
vectors). The other equations contain a gauge structure and the
scalar ``quantum potential'' which indicates some arbitrariness in
a theory. Formal results of this part need to be reexamined in
order to establish physical meaning and to find concrete
applications.

In the final (fourth) paper we will expand extended Schwinger's
quantization scheme on the case of superspase (considered as a quotient
space $SP_{4}/SO(1, 3)$).

The investigations performed in the present serie of works shows that
Schwinger's quantization scheme, extended for the case of manifolds
with a group structure (including a superspace), may be viewed as a
universal method of quantum theory. This method allow one to solve the
quantum mechanical problems in cases, where the canonical quantization
method became questionable one, because it requires new additional
assumptions.

In our opinion the approach introduced in the present works can be
useful for analyses models describing the particle-like solitons in a
collective coordinate formalism (\cite{quant1}, \cite{quant2},
\cite{quant3}).

\section{Variation principle in quantum mechanics}
\label{s1}

The variational principle, adapted for
the purposes of quantum mechanics,
was investigated by Schwinger in 1951 \cite{schwinger}. In these works
Schwinger has analyzed the special case of the theory characterized by
the Lagrangian with a linear kinetic part on generalized velocities
$\{\dot{q}^{\mu}:\mu=\overline{1,n} \}$ (here
$\{q^{\mu}:\mu=\overline{1,n} \}$ are generalized coordinates of the
dynamical system). He has showed that
the Heisenberg equations of motion
and commutation relations consistent with them can be obtained within
the framework of a unique scheme.

Schwinger's quantization approach is based on the assumption about
the existence of the hermitan operator of the action functional
$S[q,\dot{q}]$. Using this functional the variation of the propagator,
caused by the infinitesimal unitary transformation of the complete set
of commuting observables $\{\alpha\}$ can be defined as

\begin{equation}
\delta\left\langle \alpha_{1}, t_{1} \right|\left.\alpha_{2},
t_{2} \right\rangle=\frac{i}{\hbar}\left\langle \alpha_{1}, t_{1}
\right| \delta S[q,\dot{q}]\left| \alpha_{2}, t_{2} \right\rangle,
\label{11}
\end{equation}
where $\left| \alpha_{1,2}, t_{1,2} \right\rangle$ are initial and
final states of the dynamical system (i.e. eigenvectors of operators of
the complete set $\{\alpha\}$ for the moments of time $t_{1,2}$ ).
The variation of the action functional $S[q,\dot{q}]$ satisfies the
relation

\begin{equation}
\delta S[q,\dot{q}]=G(t_{2})-G(t_{1})
\label{12}
\end{equation}
where $G=G(t)$ is the hermitan generator of a unitary transformation.

As far as the variation of the action functional is completely
determined by
the variations of generalized coordinates $\{q^{\mu}\}$
and time $t$, the condition (\ref{12}) establishes the connection
between the infinitesimal unitary transformation in (\ref{11})
and variations $\delta q$ and $\delta t$. Variations that satisfy the
equation (\ref{12}) are called {\it permissible variations}.\footnote{
Permissible variations are called {\it elementary} (or c-number)
ones if they commute with all the operators of a model.
An elementary variation is equivalent to the unit operator.
In \cite{schwinger} Schwinger has considered the model with
elementary variations as permissible ones.}

Taking into account (\ref{11}), (\ref{12}) and the unitary nature of
permissible variations one can make a conclusion that the variation
$\delta A$ of an arbitrary operator $A$, caused by variations
$\delta q, \delta t$ is determined by the following expression

\begin{equation}
\delta A=\frac{1}{i\hbar}\left[ A, G\right]
\label{13}
\end{equation}

As far as in (\ref{13}) $\delta A$ and $G$ contain variations
$\delta q, \delta t$, this relation can be interpreted by two
ways. If the commutation relations of the model are known, one can
find the explicit form of the variations $\delta q$, $\delta t$
from (\ref{13}). On the other hand, if $\delta q$ and $\delta t$
are given, the relation (\ref{13}) can be viewed as the condition
that
determines the algebra of the commutation relations of a
model. In Schwinger's quantization scheme one use exactly the
second variant of the interpretation of the relation (\ref{13}).

Further we extend Schwinger's quantization scheme for the case of a
non-linear stationary model, in which the kinetic part of the action
depends not only on velocities $\{\dot{q}^{\mu}:\mu=\overline{1,n}\}$,
but also on coordinates $\{q^{\mu}:\mu=\overline{1,n}\}$.
This fact necessarily causes an operator ordering problem in the
Lagrangian which contains non-commutative operators $q$ and $\dot{q}$.
The Lagrangian determines the action as

\begin{equation}
S[q,\dot{q}]=\int_{t_{1}}^{t_{2}} L(q,\dot{q})\,dt
\label{14}
\end{equation}
The action functional (\ref{14}) is a starting conception in
Schwinger's scheme. To investigate the physical meaning of the
theory based on the action (\ref{14}) with a given Lagrangian one
should make some a priori assumptions about the properties of
operator variables $q$ and $\dot{q}$ (see section \ref{s3} for
further discussion).

To determine the form of the generator $G=G(t)$ of a unitary
transformation let us consider the special case of the coordinate
transformation

\begin{equation}
\overline{q}(t)=q(t)+\delta q(t)
\label{15}
\end{equation}
According to the properties of canonical transformations,
the Lagrangian can be expressed in terms of kinetic and dynamic
parts

\begin{equation}
L=L_{kin}-H
\label{16a}
\end{equation}
The variation of the kinetic part
under the transformation (\ref{15}) $L_{kin}$ satisfies the condition

\begin{equation}
\delta L_{kin}=-\frac{dK}{dt}
\label{16b}
\end{equation}
where $K=K(q,\dot{q},\delta q)$ is some homogeneous function of
$\{\delta q^{\mu}\}$. Hence

\begin{equation}
\delta L=-\frac{dK}{dt}-\delta H
\label{16c}
\end{equation}

Taking into account the operator equation (\ref{13}), we can rewrite
(\ref{16a})-(\ref{16c}) in the following form
\begin{equation}
\delta
L=-\frac{dK}{dt}-\frac{1}{i\hbar}[H, G]=-\frac{dK}{dt}+\frac{dG}{dt}
\label{17}
\end{equation}
On the other hand, the variation of $L$ under the transformation (\ref{15})
can be expressed by a standard way (by extracting a total time
derivative) as

\begin{equation}
\delta L=\frac{d}{dt}\left( \frac{\partial L}{\partial \dot{q}^{\mu}}
\star \delta q^{\mu}\right)+\frac{\delta L}{\delta q^{\mu}}
\star \delta q^{\mu}
\label{18}
\end{equation}
where the objects

\begin{equation}
\frac{\partial L}{\partial \dot{q}^{\mu}}\star\delta q^{\mu}:=
{\cal P}(q,\dot{q},\delta q)\,,\quad
\frac{\delta L}{\delta q^{\mu}}:={\cal E}(q,\dot{q},\delta q)
\label{19}
\end{equation}
denote homogeneous of $\delta q$ functions.
The symbol ``$\star$'' is used for the sake of visuality,
in the classical limit $\hbar\to0$
the combinations (\ref{19}) become the product of the classical
moment with $\delta q$ and contraction of the Euler-Lagrange equations
with $\delta q$ correspondingly
\footnote{
When $f=f(q)$ and $[q^{\mu}, \dot{q}^{\nu}]\ne 0$ one cannot
write down the variation $\delta f$ in the explicit form in a general
case. To demonstrate what the symbol ``$\star$'' means, let us consider
the simplest case $f(q)=q^{1}q^{2}q^{3}$. Then
$$
\delta f(x)=\delta q^{1}q^{2}q^{3}+q^{1}\delta q^{2} q^{3}+
q^{1}q^{2}\delta q^{3}:=\frac{\partial f}{\partial q^{\mu}}\star
\delta q^{\mu}\,,
$$

where
$$
\frac{\partial f}{\partial q^{1}}=q^{2}q^{3}\,,\quad
\frac{\partial f}{\partial q^{2}}=q^{1}q^{3}\,,\quad
\frac{\partial f}{\partial q^{3}}=q^{1}q^{2}\,.
$$
denote ``derivatives, obtained by omitting the factors $q^{1}$,
$q^{2}$, $q^{3}$ correspondingly. To write down the variation
$\delta f$ one have to insert $\delta q^{\mu}$ into
$\partial f/\partial q^{\mu}$, as we can see from these formulae.}.

Comparing (\ref{17}) with (\ref{18}) we obtain

\begin{equation}
\frac{dG}{dt}=
\frac{d}{dt}\left( K+\frac{\partial L}{\partial
\dot{q^{\mu}}}\star\delta q^{\mu} \right)+\frac{\delta L}{\delta
q^{\mu}}\star\delta q^{\mu}\,.
\label{110}
\end{equation}
Since (\ref{110}) must be agreed with (\ref{12}) we can write

\begin{equation}
G=K+\frac{\partial L}{\partial \dot{q}^{\mu}}\star\delta q^{\mu}\,,\quad
\frac{\delta L}{\delta q^{\mu}}\star\delta q^{\mu}=0\,.
\label{111}
\end{equation}
The first expression in (\ref{111}) gives the definition of a
generator $G$, the second one is related to dynamical equations for
operators $q^{\mu}$ in the Euler-Lagrange form.

Further development of a theory requires the explicit form of the
Lagrangian and investigation of
the transformation properties of operators.

\section{Transformation Properties of Operators}
\label{s2}

The construction of the quantum mechanics for the present model is
essentially based on the assumption that the coordinate operators
$\{q^{\mu}:\mu=\overline{1,n}\}$ form the complete set of commuting
observables. If the complete set is replaced to another one
consisted with the initial set (i.e. the
operators of the new set are the
functions of operators from the old one), the
physical meaning of a
theory does not change. In particular, when the complete set is
identified with coordinate operators $\{q^{\mu}\}$, such a change
of the complete set is nothing but the coordinate transformation
$q^{\mu}\to\overline{q}^{\mu}:=\overline{q}^{\mu}(q)$. All the
geometrical objects, which are functions of only $\{q^{\mu}\}$'s
commute with them.

In this paper we assume $[q^{\mu}, \dot{q}^{\nu}]$ to be a function of
only $\{q^{\mu}\}$'s. This assumption can be motivated as follows.
The model have to be invariant under an arbitrary (smooth) coordinate
transformation $q\to \overline{q}=\overline{q}(q)$, which is assosiated
with the change of the set of commuting observables. To obtain the
Lagrangian $L(q, \dot{q})$ in a new coordinate system we have to make a
substitution $q=q(q')$, $\dot{q}=\dot{q}(q, \dot{\overline{q}})$, where
$\dot{\overline{q}}=d[\overline{q}(q)]/dt$. The
important
expression $\dot{\overline{q}}^{\mu}=\dot{q^{\mu}}\circ \partial_{\alpha}
\overline{q}^{\mu}(q)$ holds if and only if $[q^{\mu}, q^{\nu}]=0$
and $[q^{\mu}, \dot{q}^{\nu}]=\{\mbox{some function of $q$}\}$
(see appendix).

Therefore, this assumption provides us to analyze a general situation,
where the form of the metric tensor and coordinate transformation is
not detalized. Of course, after determinating the commutation
relations, one have to verify its correspondence with the basic
assumptions.

The transformation rule of a geometrical object under a point
coordinate transformation
$q^{\mu}\to\overline{q}^{\mu}:=\overline{q}^{\mu}(q)$
depends on its operator properties and inner structure. In this
section we extend the classical definitions of a scalar, a vector and
a tensor for the case of non-commutative operators.

A {\it quantum scalar} we treat as a quantum geometrical object,
which doesn't change its value under the coordinate transformation
\begin{equation}
\overline{f}(\overline{q})=f(q)
\label{21}
\end{equation}
(here the argument in brackets points to the method of description,
not to the functional dependence in general).

A {\it quantum vector} is a quantum geometrical object with one
index which transforms under the coordinate transformation by one
of the following rules:

1. $A^{\mu}$ ($A_{\mu}$) is a {\it left-side vector}, if
\begin{equation}
\overline{A^{\mu}}(\overline{q})=\overline{a}^{\mu}_{\nu}(q)
A^{\nu}(q)\,, \quad
\left(\overline{A_{\mu}}(\overline{q})={a}_{\mu}^{\nu}(q)
A_{\nu}(q)\,\right); \label{22}
\end{equation}

2. $A^{\mu}$ ($A_{\mu}$) is a {\it right-side vector}, if
\begin{equation}
\overline{A^{\mu}}(\overline{q})=
A^{\nu}(q)\,\overline{a}^{\mu}_{\nu}(q)\,, \quad
\left(\overline{A_{\mu}}(\overline{q})=
A_{\nu}(q)\,{a}_{\mu}^{\nu}(q)\, \right); \label{23}
\end{equation}

3. $A^{\mu}$ ($A_{\mu}$) is a {\it two-side vector}, if it is a
left- and right-side vector simultaneously
\begin{equation}
\overline{A}^{\mu}(\overline{q})=A^{\nu}(q)\,\overline{a}^{\mu}_{\nu}(q)=
\overline{a}^{\mu}_{\nu}(q)A^{\nu}(q), \quad
\overline{A}_{\mu}(\overline{q})=A_{\nu}(q)\,{a}_{\mu}^{\nu}(q)=
{a}_{\mu}^{\nu}(q)A_{\nu}(q) \label{24}
\end{equation}

4. $A^{\mu}$ ($A_{\mu}$) is a {\it symmetrized vector}, if
\begin{equation}
\overline{A^{\mu}}(\overline{q})=\overline{a}^{\mu}_{\nu}(q)\circ
A^{\nu}(q)\,, \quad \left(
\overline{A_{\mu}}(\overline{q})={a}_{\mu}^{\nu}(q)\circ
A_{\nu}(q)\, \right) \label{25}
\end{equation}
with the following notations
\begin{equation}
\overline{a}^{\mu}_{\nu}=\frac{\partial \overline{q}^{\mu}}
{\partial q^{\nu}}\,,
\quad
a^{\mu}_{\nu}=\frac{\partial q^{\mu}}{\partial \overline{q}^{\nu}}
\label{26}
\end{equation}
for transformation matrices. Here the symbol ``$\circ$'' denotes the
symmetrized Jordan product \footnote{This product is defined as
$a\circ b:=\frac{1}{2}(ab+ba)$ for arbitrary operators $a$ and $b$.
See appendix for its properties}.

The transformation laws described by (\ref{22}) and (\ref{23})
correspond to non-hermitan operators. As a two-side vector we can
consider an arbitrary vector which depends on only the coordinate
operators $\{q^{\mu}\}$ (and commutes with the transformation matrices
(\ref{26})). At last the transformation laws described by
(\ref{25}) correspond to hermitan operators. For example the
operator of a generalized velocity $\dot{q}^{\mu}$ transforms as
a symmetrized vector

\begin{equation}
\dot{q}^{\mu}\to\dot{\overline{q}}^{\mu}=\overline{a}^{\mu}_{\nu}
\circ\dot{q}^{\nu}
\label{27}
\end{equation}
(we have taken into account the fact that $[q^{\mu}, \dot{q}^{\nu}]$
is a function of only $\{q^{\mu}\}$'s). Every symmetrized vector
can be expressed as the sum of left and right parts (see section
\ref{s4}).

In the remaining part of this section we summarize some useful
properties of quantum geometrical objects. Let us prove that the
contraction of symmetrized and two-side tensors is a symmetrized
tensor. We define the operator

\begin{equation}
p_{\mu}:=g_{\mu\nu}\circ \dot{q}^{\nu},
\label{28}
\end{equation}
where $g_{\mu\nu}=g_{\nu\mu}$ is a two-side tensor. In new coordinates
this operator receives the form

\begin{eqnarray}
\overline{p}_{\mu}&=&\overline{g}_{\mu\nu}\circ \dot{\overline{q}}^{\nu}
=\overline{g}_{\mu\nu}\circ\left( \overline{a}^{\nu}_{\alpha}
\circ \dot{q}^{\alpha} \right)=
\left( \overline{g}_{\mu\nu}\circ\overline{a}^{\nu}_{\alpha} \right)
\circ\dot{q}^{\alpha}\nonumber\\
&=&\left( g_{\alpha\nu}\circ a^{\nu}_{\mu}\right)\circ\dot{q}^{\alpha}
=a^{\nu}_{\alpha}\circ\left( g_{\alpha\nu}\circ \dot{q}^{\alpha} \right)
=a^{\nu}_{\mu}\circ p_{\nu}.
\label{29}
\end{eqnarray}
By a similar way one can demonstrate that

\begin{equation}
\dot{q}^{\mu}=g^{\mu\nu}\circ p_{\nu}\to \dot{\overline{q}}^{\mu}=
\overline{a}^{\mu}_{\nu}\circ \dot{q}^{\nu}.
\label{210}
\end{equation}

When $\{g_{\mu\nu}\}$ and $\{g^{\mu\nu}\}$ have the sense of covariant
and contravariant metric tensors respectively, (\ref{28})-(\ref{210})
can be interpreted as a rule for lowering and raising of the index in
quantum (symmetrized) tensors.

Similarly, the Jordan contraction (or a scalar product)
of two-side and symmetrized vectors is a quantum scalar. To prove this,
let us consider the symmetrized vector $p_{\mu}$ and the two-side
vector $v^{\mu}$. Then

\begin{equation}
\overline{p}_{\mu}\circ\overline{v}^{\mu}=
\overline{v}^{\mu}\circ\left( a^{\alpha}_{\mu}\circ p_{\alpha} \right)=
p_{\alpha}\circ \left(v^{\mu}\circ a^{\alpha}_{\mu}  \right)=
p_{\alpha}\circ v^{\alpha}.
\label{211}
\end{equation}

As to the Jordan contraction of symmetrized vectors, its properties are
more complicated than previously mentioned. Generally, such objects are
not quantum tensors and to determine them
one has to use the explicit form of commutation relations.

Now we consider the transformation properties of commutation relations:

\begin{enumerate}
\item
for a commutator between a quantum scalar $f=:f(q)$ and a quantum
vector $p_{\mu}$ we have $$ \left[ \overline{f},
\overline{p}_{\mu} \right]= \left[ f, a^{n}_{\mu}\circ p_{\nu}
\right]= a^{n}_{\mu}\circ \left[ f, p_{\nu} \right] $$ where
$[f,a^{\mu}_{\nu}]=0$ is assumed. According to
(\ref{21})-(\ref{25}), this object is a quantum vector;

\item
a commutator between a two-side symmetrized vectors and a quantum
vector $p_{\mu}$ reads

\begin{eqnarray}
\left[ \overline{v}^{\mu},\overline{p}_{\nu} \right]&=&
\left[ \overline{a}^{\mu}_{\alpha} v^{\alpha}, a^{\beta}_{\nu}\circ
 p_{\beta}\right]=a^{\beta}_{\nu}\left[\overline{a}^{\mu}_{\alpha}
v^{\alpha}, p_{\beta}\right] \nonumber \\
&=&a^{\beta}_{\nu}\circ
\left( \overline{a}^{\mu}_{\alpha}\circ [v^{\alpha}, p_{\beta}]+
v^{\alpha}\circ [\overline{a}^{\mu}_{\alpha}, p_{\beta}]\right)\nonumber
\end{eqnarray}
The second term in the right hand side does not have a tensor sense,
therefore $[v^{\mu}, p_{\nu}]$ is not a quantum tensor;

\item
for a commutator between two symmetrized vectors we write

\begin{eqnarray}
\left[ \overline{A}^{\mu}, \overline{B}^{\nu} \right]&=&
\left[ \overline{a}^{\mu}_{\alpha}\circ A^{\alpha},
\overline{a}^{\nu}_{\beta}\circ B^{\beta} \right]\nonumber \\
&=&\overline{a}^{\mu}_{\alpha}\,\overline{a}^{\nu}_{\beta}\circ
[A^{\alpha}, B^{\beta}]+\overline{a}^{\mu}_{\alpha}\circ
\left( \left[ A^{\alpha}, \overline{a}^{\nu}_{\beta} \right]\circ
B^{\beta}\right)+
\left( \overline{a}^{\nu}_{\beta}\circ
\left[ \overline{a}^{\mu}_{\alpha}, B^{\beta} \right] \right)\circ
A^{\alpha}\nonumber
\end{eqnarray}
Analogously to the previous, $[A^{\mu}, B^{\nu}]$ fails to be a quantum
tensor due to the presence of two non-covariant tensor terms in the
right hand side.
\end{enumerate}

In connection with the previously considered transformation properties of
quantum geometrical objects one can observe that there are some
fundamental complications in the formulation of quantum mechanics on a
Riemannian manifold.

Namely, the naive definition of the quantum Lagrangian for a free
particle in a curved space based on the classical expression

\begin{equation}
L_{o}=\frac{1}{2}\dot{q}^{\mu}g_{\mu\nu}(q)\,\dot{q}^{\nu}
\label{212}
\end{equation}
leads one to a conclusion that the Lagrangian, being a hermitan
operator, fails to be a quantum scalar (in the meaning of the definition
(\ref{21})), i.e. $\overline{L}_{o}(\overline{q})\ne L_{o}(q)$. In
the explicit form we write

\begin{equation}
\overline{L}_{o}=\frac{1}{2}\dot{\overline{q}}^{\mu}\overline
g_{\mu\nu} \dot{\overline{q}}^{\nu}=L_{o}+ \frac{1}{4}\left[
\dot{q}^{\alpha}, g_{\alpha\beta}a^{\beta}_{\mu}
b^{\mu}\right]-\frac{1}{8}a^{\alpha}_{\mu}\,a^{\beta}_{\nu}\,
g_{\alpha\beta}b^{\mu} b^{\nu} \label{213}
\end{equation}
where
$b^{\mu}:=b^{\mu}(q):=[a^{\mu}_{\alpha}(q),\dot{q}^{\alpha}]$.
Therefore, the main requirement for the Lagrangian (a scalar
invariance) is violated. Note that in a classical limit second
and third terms in the right hand side of (\ref{213}) vanish,
being functions of the second (or higher) order of $\hbar$.

\section{Quantum Lagrangian for Free Particle in Curved Space}
\label{s3}

Now let us concentrate our attention on the formulation of the quantum
mechanics for a free particle in a Riemannian space equipped with the
metric ${g_{\mu\nu}(q)}$. The starting step in this formulation
consists of construction of the quantum Lagrangian which is invariant
under the coordinate transformation
$q\to\overline{q}=\overline{q}(q)$.
We introduce the following form of the Lagrangian

\begin{equation}
L=\frac{1}{2}\dot{q}^{\mu}g_{\mu\nu}(q)\,\dot{q}^{\nu}-U_{q}(q)
\label{31}
\end{equation}
where $U_{q}=U_{q}(q)$ is some function which permits the
function (\ref{31}) to be a quantum scalar under a point
coordinate transformation ($U_{q}$ may be conditionally called as
``quantum potential'' \footnote{The term ``quantum potential''
is used in several works in
somewhat different meaning \cite{sugano}.
Therefore we use this term in quotation marks.}). Its explicit
form is unknown at this stage, because
to determine it we have to use commutation relations.

For the operators of quantum mechanics described by (\ref{31}) we
make the following basic assumptions

\begin{enumerate}

\item
$[q^{\mu}, q^{\nu}]=0$ and $\{q^{\mu}:\mu=\overline{1,n}\}$
form the complete set of commuting observables;

\item
$[q^{\mu}, \dot{q}^{\nu}]$ is a function of only $\{q^{\mu}\}$'s.
\end{enumerate}
These assumptions permit us to conclude that $U_{q}$ is a function of
only $\{q^{\mu}\}$'s.

Taking the total time derivative of $[q^{\mu}, q^{\nu}]=0$ we can see that
$$
\frac{1}{i\hbar}[q^{\mu}, \dot{q}^{\nu}]=f^{\mu\nu}(q)=f^{\nu\mu}(q)\,.
$$
Note that the ordering of factors in (\ref{31}), which satisfies the hermitan
condition is not unique. We con take the expression
$\dot{q}^{\mu}\circ(g_{\mu\nu}\circ\dot{q}^{\nu})$ instead of
$\dot{q}^{\mu}g_{\mu\nu}\dot{q}^{\nu}$. Such a replacement leads to another
function $U_{q}(q)$. The scalar Lagrangian $L$ is the same (one can
demonstrate this fact after determination the commutation relations).

Now we consider the transformation properties of $U_{q}(q)$ and
$L(q, \dot{q})$under the infinitesimal coordinate transformation
$q^{\mu}\to q^{\mu}+\delta q^{\mu}(q)$. Taking into account the
fact that under such a transformation the velocity operator and the metric
change as

$$
\delta_{c}\dot{q}^{\mu}=\dot{q}^{\alpha}\circ\partial_{\alpha}
\delta q^{\mu},,\quad
\delta_{c}g_{\mu\nu}=-g_{\mu\alpha}\partial_{\nu}\delta q^{\alpha}-
g_{\alpha\nu}\partial_{\mu}\delta q^{\alpha}\,,
$$
we can write

\begin{equation}
\delta_{c}L_{o}=\frac{1}{4}
\left[ \left[ \partial_{\alpha}\delta q^{\mu},
\dot{q}^{\alpha}\right]g_{\mu\nu}, \dot{q}^{\nu} \right]\,,
\label{d1}
\end{equation}
where we have used the basic assumptions.

Because the Lagrangian $L=L_{o}-U_{q}$ is a scalar, we have $\delta_{c}L=0$,
then

\begin{equation}
\delta_{c}U_{q}=
\delta_{c}L_{o}=\frac{1}{4}
\left[ \left[ \partial_{\alpha}\delta q^{\mu},
\dot{q}^{\alpha}\right]g_{\mu\nu}, \dot{q}^{\nu} \right]\,,
\label{d2}
\end{equation}

This formula shows, that neither $L_{o}$ nor $U_{q}$ is a quantum scalar,
this property has the combination $L=L_{o}-U_{q}$.

Further, let us write down the variation of the operator $L(q, \dot{q})$
under the alteration of its arguments:

\begin{equation}
\delta L(q, \dot{q}):=
L(q+\delta q, \dot{q}+\delta \dot{q})-L(q, \dot{q})\,.
\label{d3}
\end{equation}
If the variation ``$\delta$'' satisfies

$$
\frac{d}{d\,t}\delta q^{\mu}=\delta \dot{q^{\mu}}
$$
and

$$
\delta g_{\mu\nu}=g_{\mu\nu}(q+\delta q)-g_{\mu\nu}(q)=\delta q^{\alpha}
\partial_{\alpha} g_{\mu\nu}\,,
$$
we obtain

\begin{equation}
\delta L(q, \dot{q})=\frac{1}{2}\dot{q}^{\mu}
(\delta q^{\alpha}\partial_{\alpha} g_{\mu\nu}+g_{\mu\alpha}\partial_{\nu}
\delta q^{\alpha}+g\_{\alpha\nu}\partial_{\mu}\delta q^{\alpha}) q^{\nu}+
+\frac{1}{4}\left[ \left[ \partial_{\alpha}\delta q^{\mu}, \dot{q}^{\alpha}
\right] g_{\mu\nu}, \dot{q}^{\nu} \right]-\delta q^{\alpha}
\partial_{\alpha} U_{q}(q)\,.
\label{d4}
\end{equation}

The Lie variation of the geometrical object $F(q)$ under the transformation
$q\to q+\delta q$ is defined as

$$
\delta_{L}F(q)=\delta_{c} F(q)-\delta q^{\mu}\partial_{\mu}F(q)\,.
$$
In particular, the Lie variation of the non-scalar function $U_{q}(q)$
has the form

$$
\delta_{L}U_{q}=
\frac{1}{4}\left[ \left[ \partial_{\alpha}\delta q^{\mu}, \dot{q}^{\alpha}
\right] g_{\mu\nu}, \dot{q}^{\nu} \right]-\delta q^{\alpha}
\partial_{\alpha} U_{q}(q)\,.
$$

Hence we can rewrite (\ref{d4}) as

\begin{equation}
\delta L=-\frac{1}{2}\dot{q}^{\mu}g_{\mu\nu}\dot{q}^{\nu}+\delta_{L}U_{q}\,.
\label{d5}
\end{equation}

The variation $\delta(\dots)$ is permissible, if $\delta L$ reduces
to the total time derivation of some function (see section (\ref{s1})).
Without loss the generality we can assume that $\delta L$.

Comparing the factors corresponding to different powers of $\dot{q}$
we can find that

\begin{eqnarray}
&&\delta_L g_{\mu\nu}=0\nonumber \\ &&\delta_{L}U_{q}=\frac{1}{4}
\left[ \left[
\partial _{\alpha} \delta q^{\mu}, \dot{q}^{\alpha} \right]
g_{\mu\nu}, \dot{q}^{\nu} \right] - \delta q^{\mu}\partial _{\mu}
U_{q}=0. \label{34}
\end{eqnarray}

The first equation in (\ref{34}) means that $\{\delta q^{\mu}\}$
is a Killing vector for the metric $\{g_{\mu\nu}\}$.
Every solution of the Killing equation $\delta_{L} g_{\mu\nu}=0$
can be decomposed as $\delta q^{\mu}=\varepsilon^{a} v^{\mu}_{a}$, where
$\varepsilon^{a}=const$ (an infinitesimal $c$-numbers) and
$\{v^{\mu}_{a}:  \mu=\overline{1,n},\: a=\overline{1,m}\}$ are $m$
independent solutions of this equation.

Comparing (\ref{d5}) with the general expression of the variation of the
Lagrangian we find that
$\delta H=0$ (for unknown $H$). As a consequence, the generator
of permissible variations $\delta q^{\mu}=\varepsilon^{a} v^{\mu}_{a}$
takes the form

\begin{eqnarray}
&&G=p_{\mu}\circ\delta q^{\mu}=\left( p_{\mu}\circ v^{\mu}_{a} \right)
\varepsilon^{a}:= \varepsilon^{a} p_{a}\nonumber \\
&&p_{a}:=p_{\mu}\circ v^{\mu}_{a}\,,\quad
p_{\mu}=g_{\mu\nu}\circ \dot{q}^{\nu}
\label{36}
\end{eqnarray}
where the permissible variations
$\delta q^{\mu}=\varepsilon^{a} v_{a}^{\mu}$ are Killing
vectors expressed as the linear combination of the independent solutions
$\{v_{a}\}$ of the equations $\delta_{L} g_{\mu\nu}=0$.

Therefore we conclude that the features of quantum mechanics on a curved
space $V_{n}$ essentially depend on the properties of its group of
isometries. This group appears in a theory in the generator $G(t)$.
The group properties of $\{v^{\mu}_{a}\}$ are expressed as

\begin{equation}
v^{\mu}_{a}\partial _{\mu}v^{\nu}_{b}-
v^{\mu}_{b}\partial _{\mu}v^{\nu}_{a}=c^{c}{}_{ab}v^{\nu}_{c}
\label{37}
\end{equation}
where $c^{a}{}_{bc}$ are the structure constants of a group.
The set of vector fields $\{v^{\mu}_{a}\partial _{\mu}\}$
forms the representation of the Lie algebra induced by the
representation of the Lie group of isometries.

On the other hand, the variation $\delta L$ can be rewritten as

\begin{equation}
\delta L=\frac{d}{dt}(G)-
\dot{p}^{\mu}\circ \delta q^{\mu}+
\frac{1}{2}\dot{q}^{\mu} (\delta q^{\alpha} \partial_{\alpha}
g_{\mu\nu})\dot{q}^{\nu}-\delta q^{\mu} \partial_{\mu} U_{q}+
\frac{1}{2}[\delta q^{\mu}, \dot{g}_{\mu}]
\label{d6}
\end{equation}
where

$$
g_{\mu}=\frac{1}{2}[\dot{q}^{\nu}, g_{\mu\nu}]
$$

When the symbol $\delta(\dots)$ corresponds to
the permissible variations this equation falls into two equations:
the first one, that describes the conservation if the generator

$$
\frac{d}{d\,t}G=0\,,
$$
and the second one, that contains the equations of motion:

\begin{equation}
\dot{p}_{\mu}\circ \delta q^{\mu}=
\frac{1}{2}\dot{q}^{\mu}(\partial_{\alpha} g_{\mu\nu} \delta q^{\alpha})
\dot{q}^{\nu}-\delta q^{\mu}\partial_{\mu} U_{q}
+\frac{1}{2}[\delta q^{\mu},
\dot{g}_{\mu}]
\label{main}
\end{equation}
In order to eliminate $\delta q^{\mu}$ we have to use the canonical
commutation relations (unknown at this stage).

At the end of this section we point out the fact that
then for an arbitrary operator $A$ we can write

\begin{equation}
\delta\frac{dA}{dt}=\frac{1}{i\hbar}\left[ \frac{dA}{dt}, G
\right]= \frac{1}{i\hbar}\frac{d\left[A, G \right]}{dt}-
\frac{1}{i\hbar}\left[A, \frac{dG}{dt} \right]= \frac{d\delta
A}{dt} \label{39}
\end{equation}
i.e. total time derivation commutes with the operation ``$\delta$''
of permissible variations.

\section{Commutation Relations in Case of Simply Transitive Group}
\label{s4}

In the present paper we will consider the simplest case of quantum
mechanics on a curved space. Namely, we restrict the group of
isometries of $V_{n}$ to be a simply transitive transformation group
on $V_{n}$. Such a simplification permits commutation relations to be
directly determined from (\ref{13}).

It is important to point out here that due to $U_{q}=U_{q}(q)$
this function does not make any contribution into the generator.
Therefore the explicit form of $U_{q}$ is not required in this section.
Moreover, it can be calculated when the commutation relations are
given.

In the case of the simply transitive group acting on $V_{n}$ the
number of independent Killing vectors equals to the dimension of
$V_{n}$, i.e. $m=n$ and ${\rm Rg}(v^{\mu}_{a})=n$. Then we can
introduce the inverse of $\{v^{\mu}_{a}\}$ as

$$
e^{a}_{\mu}\,v^{\nu}_{a}=\delta^{\nu}_{\mu},\quad
e^{\alpha}_{\mu}\,v^{\mu}_{b}=\delta^{a}_{b}
$$
which obeys the Maurer-Cartan equation

$$
\partial _{\mu} e^{a}_{\nu}-\partial _{\nu} e^{a}_{\mu}=
-c^{a}{}_{bc}\,e^{b}_{\mu}\,e^{c}_{\nu}
$$

To determine commutation relations we use (\ref{13}). The symbol
$\delta$ corresponding to the permissible variations in our case
means the shift

$$ \delta F(q, \dot{q})=F(q+\delta q, \dot{q}+\delta \dot{q})-
F(q,\dot{q})\,, $$ for any operator $F(q, \dot{q})$. Such a
variation corresponds to a unitary transformation that acts on
different objects equally.

At first let us employ this relation for coordinate operators:

\begin{equation}
\delta q^{\mu}=\frac{1}{i\hbar}\left[ q^{\mu}, G \right]
\label{41}
\end{equation}
Using the explicit form of the generator of permissible variations
(\ref{36}) and $\delta q^{\mu}=\varepsilon^{a} v^{\mu}_{a}$ due to the
arbitrariness of c-number parameters $\{\varepsilon^{a}\}$ we obtain
from (\ref{41})

\begin{equation}
\left( \delta^{\mu}_{\nu}-\frac{1}{i\hbar}[q^{\mu}, p_{\nu}] \right)
\circ v^{\nu}_{a}=0
\label{42}
\end{equation}
Multiplying (\ref{42}) on the inverse matrix $\{e^{a}_{\alpha}\}$
we obtain the following commutation relation

\begin{equation}
\left[ q^{\mu}, p_{\nu} \right]=i\hbar \delta^{\mu}_{\nu}
\label{43}
\end{equation}

Further let us consider the operator equation
\begin{equation}
\delta p_{\mu}=\frac{1}{i\hbar}\left[ p_{\mu}, G \right]
\label{44}
\end{equation}

Under the transformation $q\to q+\delta q(q)$
the symmetrized vector $p_{\mu}$ changes as
$$
\delta p_{\mu}=-\frac{\partial \delta q^{\nu}}{\partial q^{\mu}}
\circ p_{\nu}
$$
By making use of this relation and (\ref{44}) one easily finds
\begin{equation}
\left[ p_{\mu}, p_{\nu} \right]\circ v^{\nu}_{a}=0
\label{45}
\end{equation}
Multiplying (\ref{45}) on the inverse matrix $\{e^{a}_{\mu}\}$
we have
\begin{equation}
\left[ p_{\mu}, p_{\nu} \right]=0
\label{46}
\end{equation}
Therefore, the commutation relations (\ref{43}) and
(\ref{46}) correspond to canonical commutation relations for the
canonical momentum $p_{\mu}$ conjugate to the coordinate $q^{\mu}$.
The other commutation relations of the theory can be calculated
using (\ref{43}) and (\ref{46}). In particular, (\ref{44}) is
equivalent to

\begin{equation}
\left[ q^{\mu}, \dot{q}^{\nu} \right]=i\hbar g^{\mu\nu}\,.
\label{47}
\end{equation}
We can observe that the basic assumption about the commutator
$[q^{\mu}, \dot{q}^{\nu}]$ is in accordance with (\ref{57}).
For an arbitrary two-side geometrical object $F=F(q)$ we have

\begin{equation}
\left[ F, p_{\mu} \right]=i\hbar \partial _{\mu} F
\label{48}
\end{equation}
For example let us consider the commutation relation between the
current operators $p_{a}:=p_{\mu}\circ v^{\mu}_{a}$ (which are
quantum scalars according to (\ref{211})). A direct calculation
using (\ref{43}), (\ref{46}) and (\ref{37}) leads to

\begin{equation}
\left[ p_{a}, p_{b} \right]=-i\hbar c^{c}{}_{ab} p_{c}
\label{49}
\end{equation}
Now we discuss the transformation properties of commutation
relations under a coordinate transformation. Obviously, the
following commutation relation can be carried out for any two-side
geometrical object $F=F^{A}(q)$, where $A=\{\alpha_{1}, \dots
\alpha_{k}\}$ is a multiindex:

\begin{equation}
\left[ F^{A}, p_{\mu} \right]=i\hbar \partial _{\mu} F^{A}
\label{410}
\end{equation}
In new coordinates $F^{A}$ and $p_{\mu}$ take the form

\begin{equation}
\overline{p}_{\mu}=a^{\nu}_{\mu}\circ p_{\nu}\,,
\quad
\overline{F}^{A}=\overline{a}^{A}_{B}\,F^{B}
\label{411}
\end{equation}
where

$$
\left[ F^{A}, a^{\mu}_{\nu} \right]=0\,,
\quad
\overline{a}^{A}_{B}=\overline{a}^{\alpha_{1}}_{\beta_{1}}
\dots\overline{a}^{\alpha_{k}}_{\beta_{k}}
$$
Then
\begin{equation}
\left[ \overline{F}^{A}, \overline{p}_{\mu} \right]=
\left[ \overline{a}^{A}_{B}\,F^{B}, a^{\nu}_{\mu}\circ p_{\nu} \right]=
i\hbar a^{\nu}_{\mu}\partial _{\nu}\left( \overline{a}^{A}_{B} F^{B} \right)
=i\hbar \overline{\partial }_{\mu}\overline{F}^{A}
\label{412}
\end{equation}

This result shows that all commutation relations are form invariant
under a general coordinate transformation. Therefore, the procedure of
determination of the commutation relations is self-consistent.

\section{Determination of Quantum Correction}
\label{s5}

To derive the quantum Lagrangian of a free particle in a curved space
it is necessary to find the quantum correction $U_{q}$.
Our procedure of its determination is based on the construction of the
invariant norm of a quantum vector.

In the case of a two-side vector $\{A^{\mu}\}$ a norm has the
following form

\begin{equation}
(A, A)=\|A\|^{2}:=A_{\mu} g^{\mu\nu} A_{\nu}=
A^{\mu} g_{\mu\nu} A^{\nu}.
\label{51}
\end{equation}

If the quantum vector $\{A^{\mu}\}$ does not commute with $q^{\mu}$,
the expression (\ref{51}) fails to be a quantum scalar according to
section \ref{s3}.

To extend the expression (\ref{51}) for the case of a symmetrized
vector let us consider the transformation law of $p_{\mu}$ under a
general coordinate transformation:

\begin{equation}
\overline{p}_{\mu}=a^{\nu}_{\mu}\circ p_{\nu}= a^{\nu}_{\mu}
p_{\nu}-\frac{i\hbar}{2}\partial _{\nu} a^{\nu}_{\mu} \label{52}
\end{equation}
or

\begin{equation}
\overline{p}_{\mu}=a^{\nu}_{\mu}\circ p_{\nu}= p_{\nu}
a^{\nu}_{\mu}+\frac{i\hbar}{2}\partial _{\nu} a^{\nu}_{\mu}
\label{53}
\end{equation}
The contracted Christoffel symbol
$\Gamma_{\mu}:=\Gamma^{\alpha}_{\mu\alpha}$ transforms as

\begin{equation}
\overline{\Gamma}_{\mu }=a^{\nu}_{\mu}\,\Gamma_{\nu}+\partial
_{\nu}a^{\nu}_{\mu}
\label{54}
\end{equation}
We define two non-hermitan operators

\begin{equation}
\pi_{\mu}:=p_{\mu}+\frac{i\hbar}{2}\Gamma_{\mu}\,,
\quad
\pi_{\mu}^{\dag}:=p_{\mu}-\frac{i\hbar}{2}\Gamma_{\mu}
\label{55}
\end{equation}
with the following properties:

$$
(\pi_{\mu})^{\dag}=\pi_{\mu}^{\dag}\,,\quad
(\pi_{\mu}^{\dag})^{\dag}=\pi_{\mu}\,,\quad
p_{\mu}=\frac{1}{2}(\pi_{\mu}+\pi_{\mu}^{\dag})
$$
\begin{equation}
[\pi_{\mu}, \pi_{\nu}]=0\,,\quad
[\pi_{\mu}^{\dag}, \pi_{\nu}^{\dag}]=0\,,\quad
[\pi_{\mu}, \pi_{\nu}^{\dag}]=i\hbar \partial _{\mu}\Gamma_{\nu}=
i\hbar \partial _{\mu}\Gamma_{\nu}
\label{56}
\end{equation}
Using (\ref{54})-(\ref{56}) one can observe that
$\pi_{\mu}$ and $\pi_{\mu}^{\dag}$ behave under the coordinate transformation
as left-side and right-side quantum vectors correspondingly:

\begin{equation}
\overline{\pi}_{\mu}=a^{\nu}_{\mu}\pi_{\nu}\,,\quad
\overline{\pi}_{\mu}^{\dag}=\pi_{\mu}^{\dag}a^{\nu}_{\mu}.
\label{57}
\end{equation}
Taking into account the transformation laws of $\pi_{\mu}$ and
$\pi_{\mu}^{\dag}$ we introduce the quantum norm of the symmetrized
vector $p_{\mu}$ as

\begin{equation}
(p, p)=\|p\|^{2}:=\pi_{\mu}^{\dag} g^{\mu\nu} \pi_{\nu}
\label{58}
\end{equation}
Similarly, one can define two non-hermitan operators connected
with $\dot{q}^{\mu}$:

\begin{equation}
V^{\mu}:=\dot{q}^{\mu}-\frac{i\hbar}{2} \Phi ^{\mu}\,,\quad
V^{\mu\dag}:=\dot{q}^{\mu}+\frac{i\hbar}{2} \Phi ^{\mu}
\label{59}
\end{equation}
with the following transformation properties
\begin{equation}
\overline{V}^{\mu}=\overline{a}^{\mu}_{\nu}\,V^{\nu}\,,\quad
\overline{V}^{\mu\dag}=V^{\nu\dag}\,\overline{a}^{\mu}_{\nu}\,,\quad
\label{510}
\end{equation}
where $ \Phi^{\mu}:=g^{\alpha\beta}\Gamma^{\mu}{}_{\alpha\beta}$.

Due to (\ref{510}) we introduce the quantum norm of the symmetrized
vector $\dot{q}^{\mu}$ as

\begin{equation}
(\dot{q}, \dot{q})=\|\dot{q}\|^{2}=V^{\mu\dag} g_{\mu\nu} V^{\nu}.
\label{511}
\end{equation}

Further, taking into account the connection
$p_{\mu}=g_{\mu\nu}\circ \dot{q}^{\nu}$ one can directly prove

\begin{equation}
(p, p)=(\dot{q}, \dot{q})
\label{512}
\end{equation}
i.e. introduced above quantum norms of velocity and the momentum
operators have the same value, as it must be.

These properties lead the Lagrangian to be written in the following
form

\begin{equation}
L=\frac{1}{2}(\dot{q}, \dot{q})
\label{513}
\end{equation}
Rewriting this relation using (\ref{59}) and (\ref{510}) we find the
explicit form of $U_{q}$

\begin{equation}
U_{q}=-\frac{\hbar^{2}}{4}\left( \partial _{\mu} \Gamma^{\mu}+
\frac{1}{2} \Gamma_{\mu} \Gamma^{\mu} \right)-
\frac{\hbar^{2}}{4}\left( \partial _{\mu} \Theta^{\mu}-
\frac{1}{2} \Theta_{\mu} \Theta^{\mu} \right) \label{514}
\end{equation}
where

\begin{equation}
\Gamma^{\mu}=g^{\mu\nu}\Gamma_{\nu}\,,\quad \Theta
^{\mu}:=\partial _{\nu} g^{\mu\nu}\,,\quad \Theta_{\mu}=g_{\mu\nu}
\Theta^{\nu} \label{515}
\end{equation}
and $$ \Gamma^{\mu}+\Theta^{\mu}+\Phi^{\mu}=0. $$ We also can
write (\ref{514}) as

\begin{equation}
U_{q}=\frac{\hbar^2}{4}\left( \partial_{\mu} \Phi^{\mu}+
\frac{1}{2} g_{\mu\nu} \Phi^{\mu}\Phi^{\nu}+ \Gamma_{\mu}
\Phi^{\mu} \right). \label{516}
\end{equation}

Using commutation relations it is easy to derive the useful identity

\begin{equation}
p_{\mu} g^{\mu\nu} p_{\nu}= \dot{q}^{\alpha} g_{\alpha\beta}
\dot{q}^{\beta}+ \frac{\hbar^{2}}{2}\left( \partial _{\mu}
\Theta^{\mu}- \frac{1}{2} \Theta_{\mu} \Theta^{\mu} \right)
\label{517}
\end{equation}
Taking into account (\ref{517}) one can rewrite (\ref{513}) as

\begin{equation}
L=\frac{1}{2} p_{\mu} g^{\mu\nu} p_{\nu}+
\frac{\hbar^{2}}{4} \left( \partial_{\mu}\Gamma^{\mu}+
\frac{1}{2}\Gamma_{\mu}\Gamma^{\mu}  \right).
\label{518}
\end{equation}

The form of the Lagrangian (\ref{518}) as far as the expression for the
norm of the quantum vector $\dot{q}$ is fixed by the commutation
relations. The choice of the {\it definitions} (\ref{511}) and (\ref{518})
is motivated by their similarity to classical expressions that contain
geometrical objects, derived from the metric $g_{\mu\nu}$ and
operators $q$, $\dot{q}$.

In order to construct the Hamiltonian of a free particle in a curved
space we consider the quantum version of the Legendre transformation

\begin{equation}
H=p_{\mu}\star \dot{q}^{\mu}-L
\label{519}
\end{equation}
where

\begin{equation}
p_{\mu}\star \dot{q}^{\mu}:=\frac{1}{2}\left( \pi^{\dag}_{\mu} V^{\mu}+
V^{\mu\dag} \pi_{\mu} \right)=(p, p)
\label{520}
\end{equation}
is a scalar.

From (\ref{519}) and (\ref{520}) we directly obtain

\begin{equation}
H=(p, p)-L=\frac{1}{2}(p, p)=L
\label{521}
\end{equation}
So, our Lagrangian is purely kinematic.

The Hamiltonian can be rewritten in another form which is similar to
a classical one:

\begin{equation}
H=p_{\mu}\circ \dot{q}^{\mu}-L-Z
\label{522a}
\end{equation}
where $Z$ is an auxiliary variable has been introduced by Sugano
\cite{sugano}

\begin{equation}
Z=-\frac{\hbar^{2}}{4} R+\frac{\hbar^{2}}{4} g^{\alpha\beta}
\Gamma^{\mu}{}_{\nu\alpha} \Gamma^{\nu}{}_{\mu\beta}.
\label{522b}
\end{equation}
Here $R$ is the scalar curvature of $V_{n}$.

As a result, we have defined all the objects appearing in quantum
mechanics with the simply transitive transformation group of isometries.

\section{Equations of Motion}

Now we rewrite the form of the Euler-Lagrange equations obtained in
section \ref{s4} (see (\ref{main})) as the following variational
equation

\begin{equation}
\delta L=\frac{d}{dt}(p_{\mu}\circ \delta q^{\mu})-
\dot{p}^{\mu}\circ \delta q^{\mu}+
\frac{1}{2}\dot{q}^{\mu} (\delta q^{\alpha} \partial_{\alpha}
g_{\mu\nu})\dot{q}^{\nu}-\delta q^{\mu} \partial_{\mu} U_{q}+
\frac{1}{2}[\delta q^{\mu}, \dot{g}_{\mu}]
\label{61}
\end{equation}
where

$$
g_{\mu}=\frac{1}{2}[\dot{q}^{\nu}, g_{\mu\nu}]
$$

From (\ref{61}) one can draw a conclusion that the generator of
canonical variations is $G=p_{\mu}\circ \delta q^{\mu}$ and the equations
of motion are contained in the relation

\begin{equation}
\dot{p}_{\mu}\circ \delta q^{\mu}=
\frac{1}{2}\dot{q}^{\mu}(\partial_{\alpha} g_{\mu\nu} \delta q^{\alpha})
\dot{q}^{\nu}-\delta q^{\mu}\partial_{\mu} U_{q}
+\frac{1}{2}[\delta q^{\mu},
\dot{g}_{\mu}]
\label{62}
\end{equation}

Using the set of commutation relations obtained above one can transform
(\ref{62}) to the form

\begin{equation}
\dot{p}_{\mu}\circ \delta q^{\mu}=f_{\mu}\circ \delta q^{\mu}+
T^{\mu\nu} \delta_{L}g_{\mu\nu}
\label{63}
\end{equation}
where

\begin{equation}
f_{\mu}=-\frac{1}{2}
p_{\alpha}\partial_{\mu} g^{\alpha\beta} p_{\beta}-
\frac{\hbar^{2}}{4}\partial_{\mu} \left(
\partial_{\alpha}\Gamma^{\alpha}+\frac{1}{2}\Gamma_{\alpha}
\Gamma^{\alpha} \right)\,,
\label{64a}
\end{equation}
where $T^{\mu\nu}=T^{\mu\nu}(q)\sim \hbar ^2$ is some tensor of
second rank. The variation $\delta q^{\mu}$ is a Killing vector,
therefore the second term in (\ref{63}) vanishes due to the
condition $\delta_{L} g_{\mu\nu}=0$. Using the decomposition
$\delta q^{\mu}=\varepsilon^{a} v_{a}^{\mu}$ we can write
(\ref{63}) as

\begin{equation}
(\dot{p}_{\mu}-f_{\mu})\circ v^{\mu}_{a}=0.
\label{65}
\end{equation}
As far as the matrix $\{v^{\mu}_{a}\}$ is invertible and describes a
two-side vector, we can eliminate it from (\ref{65}) by multiplication
on its inverse $\{e^{a}_{\mu}\}$. Finally, we obtain the equation of
motion in the following form

\begin{equation}
\dot{p}_{\mu}=-\frac{1}{2}p_{\alpha} \partial_{\mu} g^{\alpha\beta}
p_{\beta}-\frac{\hbar^{2}}{4}\partial_{\mu}
\left( \partial _{\alpha} \Gamma^{\alpha}+\frac{1}{2}
\Gamma_{\alpha}\Gamma^{\alpha} \right)
\label{66}
\end{equation}
The equation of this type was considered in \cite{sugano} from the view
of canonical quantization approach.

It is easy to prove that the equation (\ref{66}) obtained from the
action principle is equivalent to the Heisenberg equations

\begin{equation}
\dot{p}_{\mu}=\frac{1}{i\hbar}[p_{\mu}, H]\,,\quad
H=\frac{1}{2}(p, p)=\frac{1}{2}(\dot{q}, \dot{q})=L.
\label{67}
\end{equation}
The generator $G=p_{\mu}\circ \delta q^{\mu}$ conserves due to
(\ref{66}), i.e. $\dot{G}=0$. This fact confirms that our quantization
scheme is self-consistent.

\section{Hamiltonian as Generator of Time Shifts}

Let us consider the coordinate transformation $q\to q+\delta q$
caused by the infinitesimal time shift $t\to t+\delta t(t)$.

\begin{eqnarray}
\delta q^{\mu} &\!=\!& \dot{q}^{\mu}\delta t\nonumber \\ \delta
\dot{q}^{\mu} &\!=\!& \frac{d}{dt} \delta q^{\mu}-
\dot{q}^{\mu}\frac{d}{dt} \delta t. \label{71}
\end{eqnarray}

Using the commutation relations for the dynamical variables
$\{q^{\mu}\}$ and $\{\dot{q}^{\mu}\}$, we obtain the following
operator properties of the variations (\ref{71})

\begin{equation}
[\dot{q}^{\mu}, \delta q^{\nu}]=[\dot{q}^{\mu}, \dot{q}^{\nu}]
\delta t=[\delta q^{\mu}, \dot{q}^{\nu}]; \label{72a}
\end{equation}

\begin{equation}
[q^{\mu}, \delta q^{\nu}]=[q^{\mu}, \dot{q}^{\nu}]\delta t= i\hbar
g^{\mu\nu} \delta t. \label{72b}
\end{equation}
The variation of the Lagrangian $L$ caused by these variations
equals to

\begin{equation}
\delta_{t}L=\delta L+L\delta t. \label{73}
\end{equation}

It is easy to rewrite (\ref{73}) as a total time
derivative without any
referring to equations of motion (i.e. by a purely algebraic way
with the use of (\ref{71})):

\begin{equation}
\delta_{t} L=\frac{d}{dt} (L \delta t). \label{74}
\end{equation}
This fact leads the variations (\ref{71}) to be permissible ones.

Further we transformate (\ref{73}) using explicitly the dynamical
equations obtained above. By  extracting a total time derivative
we can write

\begin{equation}
\delta_{t} L=\frac{d}{dt} (L \delta t)+\frac{dL}{dt}\delta t-
(\dot{p}_{\mu}-f_{\mu})\circ \delta q^{\mu}. \label{75}
\end{equation}
The last term in (\ref{75}) contains the dynamical equations
(\ref{66}). Comparing (\ref{74}) with (\ref{75}) we can observe
that

\begin{equation}
\frac{d H}{dt}\delta t=0\,,\quad H=L. \label{76}
\end{equation}
Therefore the Hamiltonian of our model is the conserved generator
of time shifts.

\section{Hilbert Space of States}

Let the coordinate operators $\{q^{\mu}\}$ form the complete set of
commutative observables. We define it's spectrum by the equation

\begin{equation}
\hat{q}^{\mu} \left| q \right\rangle=q^{\mu} \left| q \right\rangle
\label{81}
\end{equation}
(to avoid a confusion we write an operator with the hat and
c-number without it in this formula and in similar cases below).

The eigenvectors of $\{q^{\mu}\}$ are normalized by the following
condition

\begin{equation}
\left\langle q'' \right.\left| q' \right\rangle=
\frac{1}{\sqrt{g(q'')}} \delta(q''-q'):=\Delta(q''-q').
\label{82}
\end{equation}
Here $\Delta(q)$ is the $\delta$-function on $V_{n}$ conformed to
the volume element $dV=\sqrt{g(q)}dq$, $dq=dq^{1}\dots dq^{n}$.
This function has the following properties

\begin{eqnarray}
\int F(q') \Delta(q'-q) dq &\!=\!& F(q)\nonumber \\
\frac{\partial \Delta(q'-q)}{\partial q'^{\mu}}=
-\Gamma_{\mu}(q')\Delta(q'-q)&\!-\!&\frac{\partial \Delta(q'-q)}
{\partial q^{\mu}}
\label{83}
\end{eqnarray}

\begin{eqnarray}
(F(q')-F(q))\Delta(q'-q) &\!=\!& 0\nonumber \\[2mm]
(F(q')-F(q))\frac{\partial \Delta(q'-q)}{\partial q'^{\mu}}
&\!=\!& -\frac{\partial F(q')}{\partial q'^{\mu}}\Delta(q'-q)
\label{84}
\end{eqnarray}
for an arbitrary smooth function $F(q)$ on $V_{n}$.

To construct a coordinate representation associated with the complete
set $\{q^{\mu}\}$ we need the
matrix elements of acting operators. For the
coordinate operator $q^{\mu}$ we easily find

\begin{equation}
\left\langle q'' \right| q^{\mu}\left| q' \right\rangle=
q'^{\mu}\left\langle q'' \right.\left| q' \right\rangle.
\label{85}
\end{equation}

In order to calculate
the matrix element of the momentum operator let us
consider the matrix element of the commutator
$[q^{\mu}, p_{\nu}]=i\hbar \delta^{\mu}_{\nu}$:

\begin{equation}
\left\langle q'' \right|q^{\mu}p_{\nu}- p_{\nu}q^{\mu}\left| q'
\right\rangle=i\hbar\delta^{\mu}_{\nu}\Delta(q''-q'.)
\label{86}
\end{equation}
The left hand side of the equality (\ref{86}) can be rewritten as

\begin{eqnarray}
\left\langle q'' \right|[q^{\mu}, p_{\nu}]\left| q' \right\rangle &\!=\!&
\int d V'''
\left(
\left\langle q''\right|q^{\mu}\left|q'''\right\rangle
\left\langle q''' \right|p_{\nu}\left| q' \right\rangle-
\left\langle q''\right|p_{\nu}\left|q'''\right\rangle
\left\langle q''' \right|q^{\mu}\left| q' \right\rangle
\right)\nonumber \\
&=& \int d V'''
\left(
q'''^{\mu}\Delta(q''-q''')\left\langle q''' \right|p_{\nu}
\left| q' \right\rangle-
q'^{\mu}\Delta(q'''-q')\left\langle q'' \right|p_{\nu}
\left| q''' \right\rangle
\right)\nonumber \\
&=& \left( q''^{\mu}-q'^{\mu} \right)
\left\langle q'' \right|p_{\nu}\left| q' \right\rangle\nonumber.
\end{eqnarray}
So that (\ref{86}) is equivalent to

\begin{equation}
(q''^{\mu}-q'^{\mu})\left\langle q'' \right|p_{\nu}
\left| q' \right\rangle =i\hbar \delta^{\mu}_{\nu}\Delta(q''-q').
\label{87}
\end{equation}

The formula (\ref{87}) can be viewed as the equation for unknown
$\left\langle q'' \right|p_{\nu}\left| q' \right\rangle$.
Using the properties of the $\Delta$-function we obtain the solution of
(\ref{87}) in the form

\begin{equation}
\left\langle q'' \right|p_{\mu}\left| q' \right\rangle=-i\hbar
\frac{\partial }{\partial q''^{\mu}}\Delta(q''-q')+
F_{\mu}(q'')\Delta(q''-q')
\label{88}
\end{equation}
where $F_{\mu}(q)$ is some smooth function on $V_{n}$ which will be
determined later.
Its appearance in (\ref{88}) does not lead to any inner contradictions.
To observe it let us calculate the matrix element of the commutator
$[f, p_{\mu}]$ for some operator $f$ using
(\ref{85}), (\ref{88}) and (\ref{83})-(\ref{84}). We find:

\begin{equation}
\left\langle q'' \right|[f, p_{\mu}]\left| q' \right\rangle=
\left( F_{\mu}(q')-F_{\mu}(q'') \right) \left\langle q''
\right|f\left| q' \right\rangle +i\hbar\left( \frac{\partial
}{\partial q''^{\mu}}+ \frac{\partial }{\partial
q'^{\mu}}+\Gamma_{\mu}(q') \right) \left\langle q'' \right|f\left|
q' \right\rangle. \label{89}
\end{equation}

For the case $[q^{\mu}, f]=0$ we have

\begin{equation}
\left\langle q'' \right|f(q)\left| q'
\right\rangle=f(q')\Delta(q''-q'). \label{810}
\end{equation}
So that (\ref{89}) can be reduced to

\begin{equation}
\left\langle q'' \right|[f(q), p_{\mu}]\left| q' \right\rangle=
i\hbar\frac{\partial f(q')}{\partial q'^{\mu}}\Delta(q''-q')
\label{811}
\end{equation}
(using (\ref{83})-(\ref{84}) and (\ref{810})). This result is
completely agreed with the commutator

$$
[f(q), p_{\mu}]=i\hbar\frac{\partial f(q)}{\partial q^{\mu}}.
$$

Now we turn to the explicit form of the function $F_{\mu}(q)$.
To determine it we need the matrix element of the commutator
$[p_{\mu}, p_{\nu}]=0$ which can be obtained by replacing $f(q)$
by $p_{\nu}$ in (\ref{89}):
$$
\left\langle q'' \right|[p_{\mu}, p_{\nu}]\left| q' \right\rangle=
i\hbar
\left(\frac{\partial F_{\nu}(q')}{\partial q'^{\mu}}-
\frac{\partial F_{\mu}(q')}{\partial q'^{\nu}}\right)\Delta(q''-q')=0.
$$
From this equation we find

\begin{equation}
F_{\mu}=\frac{\partial F(q)}{\partial q^{\mu}} \label{812}
\end{equation}
where $F(q)$ is some scalar function.
Due to (\ref{812}) we rewrite (\ref{88}) as

\begin{equation}
\left\langle q'' \right|p_{\mu}\left| q' \right\rangle=
-i\hbar\frac{\partial \Delta(q''-q')}{\partial q''^{\mu}}+
\frac{\partial F(q'')}{\partial q''^{\mu}}\Delta(q''-q').
\label{813}
\end{equation}

The Hermitan conjugation of (\ref{813}) due to (\ref{83})-(\ref{84})
can be expressed as

\begin{equation}
\left\langle q' \right|p_{\mu}\left| q'' \right\rangle^{\ast}=
-i\hbar\frac{\partial \Delta(q''-q')}{\partial q''^{\mu}}
-i\hbar\Gamma_{\mu}(q'')\Delta(q''-q')+ \frac{\partial
F^{\ast}}{\partial q''^{\mu}}\Delta(q''-q') \label{814}
\end{equation}
The Hermitan property of $p_{\mu}$ leads to
$$
\left\langle q'' \right|p_{\mu}\left| q' \right\rangle=
\left\langle q' \right|p_{\mu}\left| q'' \right\rangle^{\ast}
$$
then

\begin{equation}
{\rm Im} \left(\frac{\partial F}{\partial q^{\mu}}\right)=
\frac{1}{2i}\frac{\partial }{\partial q^{\mu}}(F-F^{\ast})=
\frac{\hbar}{2}\,\Gamma_{\mu.} \label{815}
\end{equation}
From the definition of $\Gamma_{\mu}$ we find

$$ \Gamma_{\mu}=\frac{1}{2}\partial_{\mu}\ln g $$

Therefore we can decompose $F(q)$ into real and imagine parts

\begin{equation}
F=-\varphi-\frac{i\hbar}{4}\ln g
\label{816}
\end{equation}
where $\varphi$ is some real-valued scalar function on $V_{n}$.

Using (\ref{816}) in (\ref{88}) we finally write the matrix element of
$p_{\mu}$:

\begin{equation}
\left\langle q'' \right|p_{\mu}\left| q' \right\rangle=
-i\hbar\frac{\partial \Delta(q''-q')}{\partial q''^{\mu}}-
\left( \frac{\partial \varphi}{\partial q''^{\mu}}+
\frac{i\hbar}{2}\Gamma_{\mu}(q'')\right)\Delta(q''-q')
\label{817}
\end{equation}
depending on an arbitrary  real-valued function $\varphi(q)$.
Its appearance in (\ref{817}) does not affect on physical states
because we can eliminate $\varphi(q)$ by the unitary transformation

\begin{eqnarray}
\left| q \right\rangle &\to& U(q)\left| q \right\rangle\nonumber
\\ p_{\mu} &\to& U p_{\mu} U^{\dag}=p_{\mu}-\partial_{\mu}\varphi
\\[1mm] q^{\mu} &\to& U q^{\mu} U^{\dag}=q^{\mu}
\label{818}
\end{eqnarray}
where
$$
U(q)=\exp\left( -\frac{1}{i\hbar}\varphi(q) \right)
$$
(see \cite{dirac}). Therefore, without loss of generality we assume
$\varphi(q)=0$.

Now we construct the coordinate representation for our model.
In order to do it we represent the wave function

\begin{equation}
\psi(q)=\left\langle q \right.\left| \psi \right\rangle
\label{819}
\end{equation}
for an arbitrary state $\left| \psi \right\rangle$.

The coordinate representation of the operator $f$ is defined by the
following formula

\begin{equation}
\hat{f}\psi(q):=\left\langle q \right|f\left| \psi \right\rangle=
\int dq' \sqrt{g(q')}\left\langle q \right|f\left| q' \right\rangle
\left\langle q' \right.\left|\psi \right\rangle
\label{820}
\end{equation}
Substituting $f=q^{\mu}$ we can obtain

\begin{equation}
\hat{q}^{\mu}\psi(q)=
\int dq' \sqrt{g(q')}\left\langle q \right|q^{\mu}\left| q'
\right\rangle \left\langle q' \right.\left|\psi \right\rangle=
q'^{\mu}\psi(q).
\label{821}
\end{equation}
Similarly, making the substitution $f=p_{\mu}$ in (\ref{820})
we write

\begin{eqnarray}
\hat{p_{\mu}}\psi(q) &=&
\int dq' \sqrt{g(q')}\left\langle q \right|p_{\mu}\left| q'
\right\rangle \left\langle q' \right.
\left|\psi \right\rangle\nonumber \\
 &=& \int dq'\sqrt{g(q')}
\left( F_{\mu}(q')\Delta(q-q')-i\hbar
\frac{\partial \Delta(q-q')}{\partial q^{\mu}} \right)\psi(q)
\nonumber \\
  &=& -i\hbar\frac{\partial \psi(q)}{\partial q^{\mu}}-
\frac{i\hbar}{2}\Gamma_{\mu}(q)\psi(q)-
\frac{\partial \varphi(q)}{\partial q^{\mu}}\psi(q).
\nonumber
\end{eqnarray}
Taking $\varphi(q)=0$ we finally have

\begin{equation}
\hat{p}_{\mu}\psi(q)=-i\hbar\left(
\frac{\partial }{\partial q^{\mu}}+\frac{1}{2}\Gamma_{\mu}(q)\right)
\psi(q)
\label{822}
\end{equation}
This representation coincides with \cite{dewitt}.
Under a general coordinate transformation the object
(\ref{822}) transforms as a vector.

Hence we have found the coordinate representation for coordinate
and momentum operators

\begin{equation}
\hat{q}^{\mu}=q^{\mu}\,,\quad
\hat{p}_{\mu}= -i\hbar \left( \frac{\partial }{\partial q^{\mu}}
+\frac{1}{2}\Gamma_{\mu} \right).
\label{824}
\end{equation}

The coordinate representation for the operators $\pi_{\mu}$
and $\pi_{\mu}^{\dag}$ can be constructed by a similar way.
Their matrix elements have the form:

\begin{eqnarray}
\left\langle q'' \right|\pi_{\mu}\left| q' \right\rangle &=&
-i\hbar\frac{\partial }{\partial q''^{\mu}}\Delta(q''-q')\nonumber \\
\left\langle q'' \right|\pi_{\mu}^{\dag}\left| q' \right\rangle
&=& -i\hbar\frac{\partial }{\partial q''^{\mu}}\Delta(q''-q')-
i\hbar\Gamma_{\mu}(q')\Delta(q''-q').
\label{825}
\end{eqnarray}
Therefore,

\begin{equation}
\hat{\pi}_{\mu}=-i\hbar\frac{\partial }{\partial q^{\mu}}\,,\quad
\hat{\pi}_{\mu}^{\dag}=-i\hbar\frac{\partial }{\partial q^{\mu}}-
i\hbar\Gamma_{\mu}.
\label{826}
\end{equation}

In order to find the coordinate representation of the Hamiltonian, we
use the formula

\begin{eqnarray}
\hat{A}(\hat{B}\psi) &=& \int dV' dV''
\left\langle q \right|A\left| q' \right\rangle
\left\langle q' \right|B\left| q'' \right\rangle
\left\langle q'' \right.\left| \psi \right\rangle\nonumber \\
&=& \int dV''
\left\langle q \right|AB\left| q'' \right\rangle
\left\langle q'' \right.\left| \psi \right\rangle=
\widehat{AB}\psi.
\label{827}
\end{eqnarray}
Putting $\hat{A}=\hat{\pi}_{\mu}^{\dag}$, $\hat{B}=g^{\mu\nu}\pi_{\mu}$
into (\ref{827}) we have

\begin{eqnarray}
2\hat{H}\psi &=&
(\pi_{\mu}^{\dag}g^{\mu\nu}\pi_{\nu})^{\wedge}\psi=
\hat{\pi}_{\mu}^{\dag}\left( g^{\mu\nu}\hat{\pi}_{\nu}\psi \right)=
-\hbar^{2}\left[ \frac{\partial }{\partial q^{\mu}}
\left( g^{\mu\nu}\frac{\partial \psi}{\partial q^{\nu}} \right)+
\Gamma_{\mu}\left( g^{\mu\nu}\frac{\partial \psi}{\partial q^{\nu}}\right)
\right]\nonumber \\
 &=& -\hbar^{2}\frac{1}{\sqrt{g}}\frac{\partial }{\partial q^{\mu}}
\left( g^{\mu\nu}\frac{\partial }{\partial q^{\nu}} \right)\psi=
-\hbar^{2}\nabla_{\mu}g^{\mu\nu}\nabla_{\nu}\psi.
\label{828}
\end{eqnarray}
Here $\nabla_{\mu}$ is covariant derivative in the metric
$\{g_{\mu\nu}\}$.

Hence, the coordinate representation of the Hamiltonian

\begin{equation}
\hat{H}=-\frac{1}{2}\hbar^{2}\nabla_{\mu}g^{\mu\nu}\nabla_{\nu}\psi
\label{829}
\end{equation}
is nothing but the Laplace operator on $V_{n}$.

The Schr\"odinger equation for a free particle on $V_{n}$ reads

\begin{equation}
-\frac{\hbar^{2}}{2}
\frac{1}{\sqrt{g}}\frac{\partial }{\partial q^{\mu}}
\left( g^{\mu\nu}\frac{\partial \psi}{\partial q^{\nu}} \right)
=E\psi.
\label{830}
\end{equation}

In the end of this section we add a remark on the form of the generator
$G$. Using (\ref{827}) we directly calculate

\begin{equation}
\hat{G}=\hat{v}^{\mu}\circ \hat{p}_{\mu}=
-i\hbar v^{\mu}\frac{\partial }{\partial q^{\mu}},
\label{831}
\end{equation}
where $\{v^{\mu}\}$ is a Killing vector (note that we need the
particular form of the Killing equation, namely
$\nabla_{\mu}v^{\mu}=0$).

\section{Conclusions}

Observing obtained results, we can conclude that our
extension of Schwinger's quantization procedure  allow one to solve
the problem of the formulation of quantum mechanics on the manifold
with a group structure without assuming non-strictly motivated
assumptions.

The main features of the present work, which have a general character,
are:

\begin{enumerate}

\item
the logical motivation of the use of Killing vectors as
permissible variations in quantum mechanics on the Riemannian
space in Schwinger's approach;

\item
the method of construction of the Lagrangian which is invariant under
a general coordinate transformation;

\item
the definition of the quantum norm of velocity and momentum operators
which is invariant under a general coordinate transformation.
\end{enumerate}

Applying them, we have rigorously defined quantum mechanics on the
manifold with the simply transitive group of isometries. The theory
includes commutative relations, Lagrangian and Heisenberg equations of
motion and seems to be self-consistent.

These results, obtained within the framework of a unified quantization
approach, are in accordance with \cite{sugano} and \cite{dewitt},
where the quantum theory was developed by means of canonical
quantization methods based on some special assumptions.

In forthcoming papers we will apply our quantization procedure
to construct a quantum theory on Riemannian manifolds with
a more complicated group structure.

\section*{Appendix. Summetrrized Jordan Product}
\setcounter{equation}{0}
\renewcommand{\theequation}{A.\arabic{equation}}

When $a$, $b$ are hermitan operators, its product $a\cdot b$
is non-hermitan in  a general case. The Hermitan condition holds for
the Jordan product

\begin{equation}
a\circ b=\frac{1}{2}(ab+ba)\,.
\label{a1}
\end{equation}
From this definition one can immediately obtain

\begin{enumerate}
\item
$a\circ b=b\circ a$;
\item
$(a+b)\circ c=a\circ c+b\circ c$;
\item
$(\alpha a)\circ b=a\circ (\alpha b)=\alpha (a\circ b)$, $\alpha\in C$;
\item
$[a, b\circ c]=[a, b]\circ c+b\circ [a, c]$;
\end{enumerate}

The Jordan product is non-associative:

\begin{equation}
a\circ (b\circ c)=(a\circ b)\circ c-\frac{1}{4}[b, [a,c]]
\label{a2}
\end{equation}
Let us concentrate our attention on the combinations appearing in our
model. The basic assumptions have the form (see section (\ref{s3})):

\begin{enumerate}
\item
$\forall \mu, \nu=\overline{1, n}$, $[q^{\mu}, q^{\nu}]=0$,
\item
$\frac{1}{i\hbar}[q^{\mu}, \dot{q}^{\nu}]=f^{\mu\nu}(q)$.
\end{enumerate}
Taking the time derivative

$$
\frac{d}{dt}[q^{\mu}, q^{\nu}]=[\dot{q}^{\mu}, q^{\nu}]+
[q^{\mu}, \dot{q}^{\nu}]+0\,,
$$
we find:

\begin{equation}
f^{\mu\nu}(q)=f^{\nu\mu}(q)\,.
\label{a3}
\end{equation}

Due to these properties, the time derivative of the operator $F(q)$
can be written as

\begin{equation}
\frac{d}{dt} F(q)=\dot{q}^{\mu}\circ \frac{\partial F(q)}
{\partial q^{\mu}}
\label{a4}
\end{equation}
(if $F(q)$ is a polynomial, the proof is elementary).

It is important to note that in some cases, that are determined by the
operator properties of multipliers, the Jordan product is
associative. Looking at the formula  (\ref{a2}) we see, that
$a\circ (b\circ c)=(a\circ b)\circ c$ if $[a, b]=0$ or
$[b, [a, c]]=0$.
Taking into account the basic assumptions we can write

\begin{equation}
f_{1}(q)\circ (\dot{q}^{\mu}\circ f_{2}(q))=
(f_{1}(q)\circ \dot{q}^{\mu})\circ f_{2}(q)\,,
\label{a5}
\end{equation}

\begin{eqnarray}
&&f_{1}(q)\circ (\dot{q}^{\mu}\circ f_{2}(q))=
f_{1}(q)\circ (f_{2}(q)\circ \dot{q}^{\mu})\nonumber \\
=(f_{1}(q)\,f_{2}(q))\circ\dot{q}^{\mu}&-&
\frac{1}{4}[f_{2}(q), \mbox{some function of $q$'s}]=
(f_{1}(q)\,f_{2}(q))\circ\dot{q}^{\mu}\,.
\label{a6}
\end{eqnarray}

\end{document}